# Wordification: A New Way of Teaching English Spelling Patterns

Lexington Whalen
College of Computing and Engineering
University of South Carolina
Columbia, SC
United States of America
lawhalen@email.sc.edu

Dalton Craven
College of Computing and Engineering
University of South Carolina
Columbia, SC
United States of America
djcraven@email.sc.edu

Shashank Comandur
College of Computing and Engineering
University of South Carolina
Columbia, SC
United States of America
comandus@email.sc.edu

Nathan Bickel
College of Computing and Engineering
University of South Carolina
Columbia, SC
United States of America
nbickel@email.sc.edu

Homayoun Valafar
College of Computing and Engineering
University of South Carolina
Columbia, SC
United States of America
homayoun@cse.sc.edu

Stanley Dubinsky
Linguistics Program
College of Arts and Sciences
University of South Carolina
Columbia, SC
United States of America
dubinsk@mailbox.sc.edu

*Abstract*—*Literacy, or the ability to read and write, is a crucial indicator of success in life and greater society. It is estimated that 85% of people in juvenile delinquent systems cannot adequately read or write, that more than half of those with substance abuse issues have complications in reading or writing and that two-thirds of those who do not complete high school lack proper literacy skills [1]. Furthermore, young children who do not possess reading skills matching grade level by the fourth grade are approximately 80% likely to not catch up at all [2]. Many may believe that in a developed country such as the United States, literacy fails to be an issue; however, this is a dangerous misunderstanding. Globally an estimated 1.19 trillion dollars are lost every year due to issues in literacy; in the USA, the loss is an estimated 300 billion [3]. To put it in more shocking terms, one in five American adults still fail to comprehend basic sentences [4]. Making matters worse, the only tools available now to correct a lack of reading and writing ability are found in expensive tutoring or other programs that oftentimes fail to be able to reach the required audience. In this paper, our team puts forward a new way of teaching English spelling and word recognitions to grade school students in the United States: Wordification. Wordification is a web application designed to teach English literacy using principles of linguistics applied to the orthographic and phonological properties of words in a manner not fully utilized previously in any computer-based teaching application.*

I. BACKGROUND

America has a literacy crisis, with 1 in 5 adults struggling to read basic sentences. A National Assessment of Educational Progress report reveals only 35% of 4th and 8th graders read proficiently and only 27% of 12th graders are proficient writers. As spelling deficits interfere with literacy, effective instruction is essential. However, teachers struggle to find time for it and believe it is inadequately addressed in the classroom. Spelling and reading resources exist for struggling students but involve $50-100/hour individual tutoring. With 60-100 hours needed to advance a student one grade, tutoring is beyond the reach of most.

In most modern classrooms, the teaching of the English language has not changed much since the rise of technology; many classrooms still rely on rote memorization of vocabulary words, and most teachers do not have the preparation nor the time to make the critical adjustments necessary for teaching students at different levels of ability [5]. Just as no two students are the same, no two students will learn to read or write at the same rate; this fact is only further emphasized when dealing with classrooms of ten or twenty students, each of which have different needs.

Anecdotal evidence collected from conversations with grade schoolteachers suggests that current methods are ineffective and insufficient for about 20% of elementary school students, completely superfluous for another 20% who need little to no instruction, and only minimally and sporadically useful to the remaining 60%. It is also the case that far too much instructional time is spent (to little positive effect) on spelling demonstration, practice, and testing. It is estimated that over 30 minutes/day are typically spent on spelling activities (representing as much as 10% of all instructional time in class) – time that might better be applied to activities which are more beneficial to learning. Our goal is to develop an instructional pathway to higher literacy that is widely, if not universally, accessible.

## II. LINGUISTIC BACKGROUND

The underlying source of spelling difficulties, especially where English is concerned, lies in the extreme irregularity and non-transparency of phonetic-orthographic mappings. Ideally, a well-ordered and phonetically transparent writing system should be such that (i) each symbol always represents the same sound, (ii) each sound is always represented by the same symbol, (iii) a single symbol represents one sound, and (iv) a single sound is represented by one symbol. English, because of centuries of language change and borrowing, violates each of these principles more than does nearly any other language. To illustrate: (i) The letter "c" can represent the [k] sound if followed by the letters "a", "o", or "u" (as in *cane*, *comb*, and *cucumber*) or the [s] sound if followed by the letters "e" or "i" (as in *cent* and *cinch*), violating principle (i). The [k] sound can be represented by the letters "c" (*cane*), "k" (*kite*), or "ck" (*luck*), violating principle (ii). The letter "x" represents the two sounds [k] and [s], as in *tax*, violating principle (iii).

The [f] sound is often represented by the digraph "ph" (*phone*), violating principle (iv). This small sample of inconsistent mapping between letters and sounds infects the entire system and makes it onerous to learn. At the same time, it should be noted that the inconsistencies themselves are not arbitrary and are explicable when grapheme-to- phoneme-to-grapheme linguistic principles are applied.

Student spelling errors, likewise, tend to collocate in relation to these principles, and it is for this reason that computerized spelling instruction (wherein the user can be instructed utilizing the principles) is a dramatic improvement over instruction that relies on word lists and memorization. This is especially true for students who are speakers of ethnic or regional dialects – e.g., African American English (AAE) or Southern White English (SWE) – or who are non-native speakers of English (e.g., Spanish-speaking children learning English at school). It has been shown that "students with good phonological awareness are in a great position to become good readers, while students with poor phonological awareness almost always struggle in reading" [6].

We take dialect variation as key to developing instructional strategies that support racially and socioeconomically marginalized learners, because most instruction and assessment approaches are designed for General American English (GAE) speakers ([7]) and there is evidence that students' dialect backgrounds influence their spelling ([8]). Prior research suggests that explicit instruction in dialect differences may be critical to maximize spelling and literacy development among nonmainstream dialect speakers ([9], [10], [11]). Accordingly, to address the need for dialect responsiveness, *Wordification* will be designed to include explicit instruction using variable pronunciations across dialects, focusing on variation across GAE, AAE, and SWE. Additionally, addressing the fact that ESL learners struggle learning words in context, as "they frequently learn dictionary forms of words, and so don't recognize them in speech when they change form or pronunciation" ([12]). *Wordification* will include, for those learners, contrasting careful and casual pronunciations, which should help to increase ESL learners' receptive vocabulary, along with improvements in their spelling. We therefore expect that students will make large gains in spelling ability by learning the spelling of words while also being shown the phoneme-grapheme patterns of the words presented to them, both in GAE and in their own dialect.

## III. TECHNICAL DESIGN

Wordification is an ambitious project. It strives to make use of the most popular architectures and paradigms to allow for fast and efficient development in the present, and simple and sustainable maintenance for the future. We have developed Wordification as a full-stack application using a combination of Next.js, GraphQL, Prisma, and TypeScript, providing a robust foundation for the project.

Wordification's internal database stores over 850 word-objects, with each word containing their respective phonemes, graphemes, example sentences and audio, school grade level, and more. Wordification also has several user types, such as student, teacher, administrator, developer, or system administrator. When a student logs in and starts a new game, that game is unique to the student who is playing and automatically records all progress made throughout. The database, in versions to be developed, will be enhanced with dialect pronunciations and casual-careful speech contrasts, to enable specialized instruction for non-native speakers and speakers of ethnic and regional dialects. All games in Wordification are pausable, and students can be registered under teachers or schools so that overall progress can be assessed; the goal is that the role of teachers in spelling will become minimal, and that teacher time can be used for other topics or to help students with severe needs.

Wordification's front end was previously written in the React library [13]. React is a JavaScript Framework maintained by Meta (formerly Facebook) [14]; Next.js is essentially React but with some additional features. Both frameworks operate on the same core principles of using components (set sections of code that operate in a specific way) to build complex front ends that are easy to maintain.

The backend was previously written in Python's Django framework, another popular framework for web applications, which uses the RESTful architecture for building web APIs [15]. An API, or application programming interface, is essentially just a grouping of established rules that allow for communication between different applications [16]. According to IBM, a REST structure is characterized by five major components: a *uniform interface* (requests for a same resource should be consistently the same), *client-server decoupling* (the client and server operate entirely independent of one another), *statelessness* (all requests must contain all data necessary to properly handle the request, there is no reliance on past information), *cacheability* (the ability to store resources on the server or client side), and a *layered system architecture* (responses and calls do not in general go through the same layers, but rather travel across different ones) [17].

Recently, we have migrated our backend framework from Django to Redwood, a full stack framework, due to several reasons related to the streamlining of future development, including improved querying, data handling, and deployment capabilities.

One significant advantage of Redwood is its seamless integration with GraphQL. GraphQL is a powerful query language that allows for easy, human-readable querying [18]. Unlike traditional REST APIs that often require multiple requests for multiple data, GraphQL enables us to retrieve multiple resources in a single request [18]. This not only reduces network overhead but also enhances the overall performance of Wordification, making it notably faster [18].

With Redwood, our backend architecture is organized in types and fields, rather than endpoints [18]. This organization provides a more structured and intuitive approach to data handling. It also prevents type errors and impossible requests, improving the safety and reliability of our application. Moreover, Redwood's integration with Prisma brings additional benefits to our development process. Prisma acts as an automated, type-safe system, simplifying database interactions and providing an intuitive and secure data access layer. Its ease of use allows our development team to focus more on implementing game logic and features rather than struggling with manual database configurations. Regarding data fetching, Redwood introduces a powerful abstraction called "cells." Cells offer a declarative approach to data fetching, enabling us to handle common scenarios such as loading screens, errors, and success states in a more streamlined manner [18]. By executing GraphQL queries and managing the lifecycle of the data, cells simplify the development process and enhance code reusability.

Redwood also offers serverless deployment capabilities by default [18]. Leveraging the benefits of serverless computing, our application can be deployed to containers managed by a cloud service provider. This approach eliminates the need to worry about operating system updates, security management, and system monitoring. Developers can focus solely on writing code, while the cloud provider takes care of resource allocation, scaling, and cost optimization. Serverless deployment ensures that Wordification is highly scalable, cost-effective, and efficiently utilizes computing resources.

In summary, our implementation of Redwood in Wordification harnesses the power of Next.js, GraphQL, Prisma, and TypeScript to deliver a robust and performant web game. Redwood's integration with GraphQL enables efficient data retrieval, while Prisma automates type-safe database interactions. The declarative nature of cells simplifies data fetching, and Redwood's serverless deployment capabilities provide scalability and cost-efficiency. With these technical design choices, Wordification offers a seamless user experience while maintaining a streamlined development process.

IV. PRODUCT

Wordification is now still a work in progress. We have designed and coded a database that includes the 850-word list of >80% of words used by elementary school students in writing, according to BSVL [19]. Words are coded for phonological, orthographic, morphological, and semantic properties, grade level, context of use, and dialect variation. Word-objects have (or will have) associated audio files: pronunciations, phonological and morphological components, illustrative sentences, and child-friendly definitions. Words are now recorded in GAE and will later be recorded in the SWE and AAE dialects. Grade level is coded according to BSVL [19]. Audio files enable speech-to-text or text-to-speech activities, and are recorded by male and female speakers, gender randomized. Dialect-specific audio will be recorded by dialect speakers.

To date, Wordification has gone through several iterations of potential applications, beginning with a solely computer version designed to be run from an executable file and now to the design mentioned in the Design section. We are presently focused on the development of two pedagogically effective game types: Word Sorting and Word Matching.

***Word sorting*** is an analytic phonics approach involving categorizing words' linguistic features. Sorting encourages students to search, compare, contrast, and analyze the features, and sorting by features helps students organize word specific knowledge and apply it to new words.

***Word building*** teaches children to assemble word components (letters, onsets and rimes, or bases and affixes, [20]). Contrary to sorting, word building is synthetic. While sorting takes words and analyzes their components, building takes components and combines them to form words.

In the Sorting game being developed, students are tasked with discriminating between two sound categories (such as between the long [o] sound represented in *rope* and the long [i] sound represented in *ripe*) – note that the database is constructed to allow the comparison of any sets of phonetic or orthographic features.

On logging in as a student, the user is presented with a choice of game type to play [Figure 1].

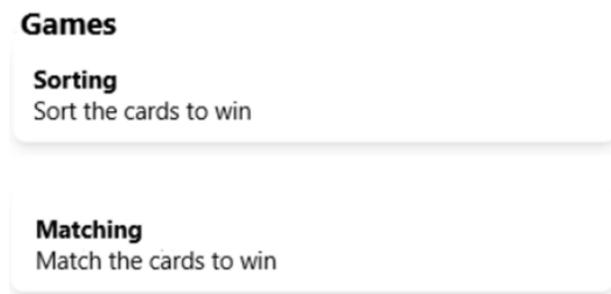

*Figure 1: Choosing between the two available games.*

Having chosen a game type (sorting in this case), the user is asked whether they wish to start a new game, resume an incomplete game, or view their completed games of this type [Figure 2].

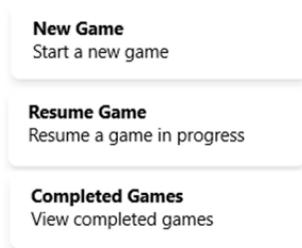

*Figure 2: Choosing to start a new game, resume a game, or view completed games.*

If a new game is selected, the user is then asked to choose which sound categories they wish to compare and how many words they wish to view in each category [Figure 3]. In this instance, having chosen to sort words having Long "I" and Long "O", the user will perform word sorting exercises involving pairings of these vowel sounds, with three spelling patterns for each vowel sound. In this case, the user will see for Long "I", words spelled with "igh", "y" and "i-consonant-e", and for Long "O", words spelled with "ow" "oa", and "o-consonant-e".

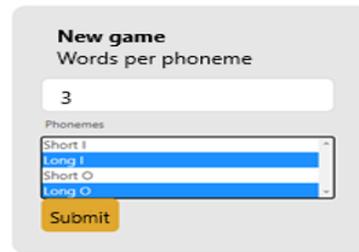

*Figure 3: Choosing between sound categories to compare and the number of words to practice on per category.*

For illustration, we can describe the intervention's presentation of the word *might*:

Game [Figure 4]: "Let's figure out the vowel sound in *might*. Here's a sentence that has *might*: *Jane might come to the party later*. The sounds that make *might* are: /m/, /aɪ/, /t/. Which is the vowel sound in *might*?" [User is presented with "Long I" and "Long O" buttons to choose from. If user clicks the correct button, the game says, "That's right. The vowel sound in *might* is /aɪ/." If the user clicks the wrong button, the game says, "That's not right. The vowel sound in *might* is not /oʊ/. Try again." [User can commit two errors before the game gives the correct answer and moves them on.]

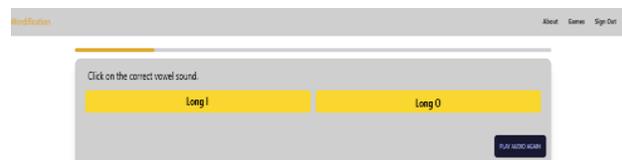

*Figure 4: Choosing between two sound categories. Note that in figures 1-3, the tested word is "might".*

Game [Figure 5]: "Now, pick the spelling pattern for the /aɪ/ in *might*." [User is presented with a choice of six grapheme buttons: "iCe", "igh", "y", "ow", "oa", and "oCe".] If user clicks the correct button, the game says, "That's right. The /aɪ/ in *might* is spelled with I-G-H." If user clicks the wrong button, the game might say, depending on their choice, "That's not right. The /aɪ/ sound in *might* is not spelled with i-consonant-e. Try again," or "That's not right, o-w is not a spelling pattern for /aɪ/. Try again." [Student can commit two errors

before the game gives correct answer and moves them on.]

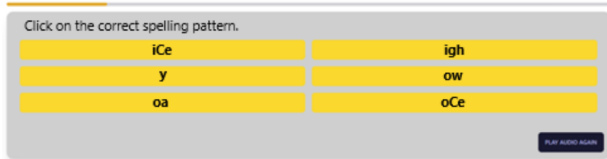

*Figure 5: Choosing the correct spelling pattern. The button turns green if it is the correct answer.*

Game [Figure 6]: "Now, in the box, type the word *might*." [User types word.] If word is spelled correctly, the game says: "That's right, *might* is spelled M-I-G-H-T. Good job." If incorrect, the game says, "That's not right. *might* has the vowel sound /ɑɪ/ and /ɑɪ/ is spelled with I-G-H in *might*." [Student can commit two errors before the game gives correct answer and moves them on.]

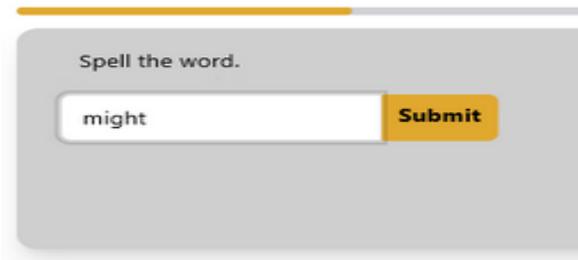

*Figure 6: Spelling the word.*

In the Matching game [Figure 7], which we will not elaborate upon here, similar phonetic and graphemic contrasts are used to build an array of word-cards, and students are tasked with flipping the cards over to find category matches.

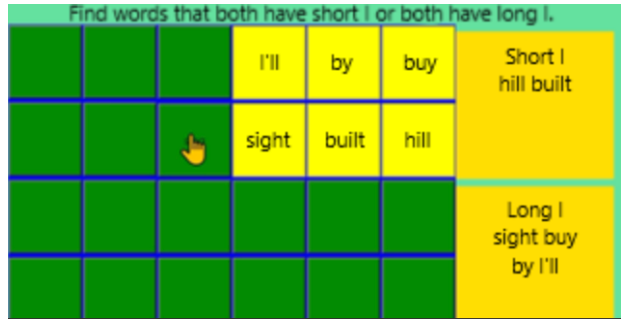

*Figure 7: Word Matching. When two cards of the same phoneme type are flipped over, they remain flipped and the color changes.*

## V. FUTURE GOALS

The future of Wordification involves three major elements: testing, expansion of the game suite, and implementation of smart grading. To begin this process, Wordification is to be tested among 1st and 2nd grade students. Initial pilot studies will involve small groups of one or two dozen students in classrooms under teacher supervision of their use and contrasting their performance and progress against students receiving standard instruction. Collecting information about its efficacy and user (student and teacher) responses, we will endeavor to improve the UI, the utility of user input, and the attractiveness of the application for the young users. Following this, we will conduct a large-scale study, involving 100-200 students, to test the usefulness of planned dialect enhancements.

The most immediate outcome of this will be a ready-to-implement classroom tool for spelling instruction. Experimenter access and instructional controls created for the testing of the application will be readily adapted for use by classroom teachers, allowing them to choose exercises and pattern drills for each student, according to that student's needs. Once *Wordification* is adopted and widely used as a tool for spelling instruction, the next stage in its development will be to internalize the application's assessment tools in order to allow individual student users to be guided by the application based on their own performance and to advance them through a spelling syllabus at a rate that is attuned to their own capacity for progress. At this stage of development, the application will no longer need teacher supervision and management (though it will continue to allow for it). By alleviating the need for hands-on, regular spelling instruction in the classroom, teachers will be able to focus on other aspects of English Language Arts, thereby allowing them to address students' needs more fully.

With *Wordification* fully gamified and equipped with the AI needed to guide students' progress individually and independently of classroom instruction, it can then be turned into a word-processing software extension. Thus, once *Wordification* is equipped to monitor a student-user's progress and guide them through a spelling syllabus, it should also be able to respond to that student's errors while typing. Accordingly, if *MS Word* is tracking a student's errors through Spell Check, it (or any other spell-checking tool) could be enabled to port the student's errors to *Wordification* so that the student would be offered spelling practice on precisely those words they misspelled on their next session in the *Wordification* game. At this point, spelling would no longer need be taught in the classroom and students (regardless of their dialect or native-speaker status) would be guided into English spelling in a manner that guarantees universal success.

VI. CONCLUSION

While technology advances at breakneck speed, the teaching of English spelling remains largely unchanged. To the extent that students are had to memorize the spellings of words without explanation of the patterns behind them, the learning of spelling and the acquisition of reading and writing skills are inhibited, leading to lower levels of literacy than what might otherwise be attained. The losses to educational attainment due to low levels of literacy are incalculable. By creating an effective and individualized method for teaching English spelling, we hope to improve students' overall literacy and contribute to their educational progress. Wordification is intended to be a paradigm-shifting invention that radically alters the learning experience, allowing for students to learn spelling patterns in a fun and efficient way.